\documentclass{stacs_proc}
\usepackage{epsf,amssymb,bbm,amsmath,exscale}
\usepackage{hyperref}

\renewcommand{\Pr}{\mbox{\bf Pr}}
\newcommand{\E}{\mbox{\bf E}}
\newcommand{\N}{\mathbbm{N}}
\newcommand{\Reals}{\mathbbm{R}}
\newcommand{\R}{\mathbbm{R}}

\renewcommand{\phi}{\varphi}
\newcommand{\Prob}{\Pr}

\begin{document}
\title{Tight Bounds for Blind Search on the Integers}
%
%
\author[ref1]{M. Dietzfelbinger}{Martin Dietzfelbinger}
\address[ref1]{{Fakult{\protect\"a}t} f{\protect\"u}r Informatik und Automatisierung, Technische Univ. Ilmenau, 98684 Ilmenau, Germany}
\email{martin.dietzfelbinger@tu-ilmenau.de}
\author[ref2]{J.E. Rowe}{Jonathan E.\ Rowe}
\address[ref2]{School of Computer Science, University of Birmingham, Birmingham B15 2TT, United Kingdom}
\email{J.E.Rowe@cs.bham.ac.uk}
%
\author[ref3]{I. Wegener}{Ingo Wegener}
\address[ref3]{FB Informatik, LS2, Universit{\protect\"a}t Dortmund, 44221 Dortmund, Germany}
\email{ingo.wegener@uni-dortmund.de}
\author[ref4]{P. Woelfel}{Philipp Woelfel}
\address[ref4]{Department of Computer Science, University of Calgary,
Calgary, Alberta T2N 1N4, Canada}
\email{woelfel@cpsc.ucalgary.ca}
\thanks{%
Work of the first author was done in part while visiting ETH
Z{\"urich}, Switzerland.  The third author was supported in part by
the DFG collaborative research project SFB 531.  Work of the last
author was done in part while at the University of Toronto, supported
by DFG grant WO1232/1-1 and by SUN Microsystems.  Joint work on this
topic was initiated during Dagstuhl Seminar  06111 on Complexity
of Boolean Functions (2006).
}
%
%
%



\begin{abstract}
  We analyze a simple random process in which a token is moved in the interval $A=\{0,\dots,n\}$:
  Fix a probability distribution $\mu$ over $\{1,\dots,n\}$. 
  Initially, the token is placed in a random position in $A$. In round $t$, a random value $d$ is chosen according to $\mu$.
  If the token is in position $a\geq d$, then it is moved to position $a-d$. 
  Otherwise it stays put.
  Let $T$ be the number of rounds until the token reaches position 0.
  We show tight bounds for the expectation of $T$ for the optimal distribution $\mu$.
  More precisely, we show that $\min_\mu\{E_\mu(T)\}=\Theta\left((\log n)^2\right)$.
  For the proof, a novel potential function argument is introduced.
  The research is motivated by the problem of approximating the minimum of a continuous function over $[0,1]$ with a ``blind'' optimization strategy.
\end{abstract}

\maketitle

\stacsheading{2008}{241-252}{Bordeaux}
\firstpageno{241}

\vskip-0.5cm
\section{Introduction}

\label{sec:problem}
For a positive integer $n$, assume a probability distribution $\mu$ on $X=\{1,\dots,n\}$ is given.
Consider the following random process. 
A token moves in $A=\{0,\dots,n\}$, as follows:
\begin{itemize}
	\item Initially, place the token in some position in $A$. 
	\item In round $t$: The token is at position $a\in A$.
	Choose an element $d$ from  $X$ at random, according to $\mu$. 
	If $d \le a$, move the token to position $a-d$ (the step is ``accepted''),
	otherwise leave it where it is (the step is ``rejected''). 
\end{itemize}
When the token has reached position 0,
no further moves are possible, and we regard the process as finished.

At the beginning the token is placed at a position
chosen uniformly at random from $\{1,\dots,n\}=A-\{0\}$. 
(For simplicity of notation, we prefer this
initial distribution over the possibly more natural uniform
distribution on $\{0,\dots,n\}$.
Of course, there is no real difference between the two 
starting conditions.) 
Let $T$ be the number of rounds needed until position $0$ is reached.
A basic performance parameter for the process is $\E_\mu(T)$. 
As $\mu$ varies, the value $\E_\mu(T)$ will vary. 
The probability distribution $\mu$ may be regarded as a strategy.
We ask: How should $\mu$ be chosen so that $\E_\mu(T)$ is as small as possible?

It is easy to exhibit 
distributions $\mu$ such that $\E_\mu(T)=O((\log n)^2)$. (All asymptotic notation in this paper refers to $n\to\infty$.)
In particular, we will see that the ``harmonic distribution''
given by 
\begin{equation}\label{eq:harmonic:distribution}
\mu_{{\rm har}}(d)=\frac1{d\cdot H_n}\,, \text{ for $1\le d \le n$,}	
\end{equation}
where $H_n=\sum_{1\le d \le n}\frac1d$ is the $n$th harmonic number,
satisfies $\E_{\mu_{\scriptstyle\rm har}}(T)=O((\log n)^2)$.
As the main result of the paper,
we will show that this upper bound is optimal up to constant factors: 
$\E_\mu(T)=\Omega((\log n)^2)$,
for every distribution $\mu$.
For the proof of this lower bound,
we introduce a novel potential function technique, 
which may be useful in other contexts. 

\vskip-0.3cm
\subsection{Motivation and Background: Blind Optimization Strategies}%
\label{sec:blind:optimization}%
Consider the problem of minimizing a function $f:[0, 1] \rightarrow \Reals$, in which the definition of $f$
is unknown: the only information we can gain about $f$ is through trying sample points. This is an instance
of a \emph{black box optimization problem}~\cite{DJW06}.
One algorithmic approach to such problems
is to start with an initial random point, and iteratively attempt to improve it by making random
perturbations. 
That is, if the current point is $x \in [0, 1]$, then we choose some distance $d \in [0, 1]$
according to some probability distribution $\mu$ on $[0,1]$, 
and move to $x+d$ or $x-d$ if this is an improvement.
The distribution $\mu$ may be regarded as a ``search strategy''.  
Such a search is ``blind'' in the sense that it does not try to estimate
how close to the minimum it is and to adapt the distribution $\mu$ accordingly. 
The problem is how to specify $\mu$.
Of course, an optimal distribution $\mu$ depends on details of the function $f$. 

The difficulty the search algorithm
faces is that 
for general functions $f$ 
there is no information about the scale of perturbations which are necessary
to get close to the minimum. 
This leads us
to the idea that the distribution might be chosen so that it is \emph{scale invariant},
meaning that steps of all ``orders of magnitude'' occur with
about the same probability. 
Such a distribution 
is described in~\cite{RH04}.
One starts by specifying a minimum perturbation size $\varepsilon$.
Then one chooses the probability density function
$h(t)=1/(pt)$ for $\varepsilon\leq t\leq 1$, and $h(t)=0$ otherwise,
where $p = \ln(1/\varepsilon)$ is the \emph{precision} of the
algorithm. 
(A random number distributed according to this density
function may be generated by taking
$d = \exp(-p u)$,
where $u$ is uniformly random in $[0, 1]$.)

For general functions $f$, no analysis of this search strategy is known, but 
 in experiments on standard benchmark functions it
 (or higher dimensional variants) 
exhibits a good performance. (For details see~\cite{RH04}.) 
From here on, we focus on the simple case where $f$ is \emph{unimodal},
meaning that it is strictly decreasing in $[0,x_0]$ and strictly increasing in $[x_0,1]$,
where $x_0$ is the unknown minimum point. 

\begin{remark}
If one is given the information that $f$ is unimodal,
one will use other, deterministic search strategies, 
which approximate the optimum up to $\varepsilon$ within 
$O(\log(1/\varepsilon))$ steps. 
As early as 1953, in~\cite{Kie53},
``Fibonacci search'' was proposed 
and analyzed, which for a given tolerance $\varepsilon$ 
uses the optimal number of steps in a very strong sense.
\end{remark} 

The ``blind search'' strategy from~\cite{RH04} can be applied to more general
functions $f$, but the following analysis is valid 
only for unimodal functions.
If the distance of
the current point $x$ from the optimum $x_0$ is $\tau\ge2\varepsilon$ then 
every distance $d$ with $\frac{\tau}{2} \le d \le \tau$
will lead to a new point with distance at most $\tau/2$.
Thus, the probability
of at least halving the distance to $x_0$ in one step is at least
\begin{math}
    \frac{1}{2}\int_{\tau/2}^{\tau} \frac{dt}{pt} = \frac{\ln 2}{2 p}, 
\end{math}
which is independent of the current state $x$.
Obviously, then, the expected number of steps before the 
distance to $x_0$ has been halved is  $2 p / \ln 2$.
We regard the algorithm to be successful if
the current point has distance smaller than $2\varepsilon$ from $x_0$. 
To reach this goal, the 
initial distance has to be halved at most $\log(1/\varepsilon)$ times,
leading to a bound of $O(\log(1/\varepsilon)^2)$ for the
expected number of steps.

The question then arises whether this is the best that can be achieved. Is
there perhaps a choice for $\mu$ that works even better on unimodal functions? 
To investigate this question, we consider a discrete version of the
situation. 
The domain of $f$ is $A=\{0,\dots,n\}$,
and $f$ is strictly increasing, so that $f$ takes its minimum at $x_0=0$.
In this case, the search process is very simple:
the actual values of $f$ are irrelevant;
going from $a$ to $a+d$ is never an improvement. 
Actually, the search process is
fully described by the simple random process from Section~\ref{sec:problem}. 
How long does it take to reach the optimal point 0,
for a $\mu$ chosen as cleverly as possible?
For $\mu=\mu_{\rm har}$, 
we will show an upper bound of
$O((\log n)^2)$,
with an argument very similar to that one leading to the
bound $O(\log(1/\varepsilon)^2)$ in the continuous case.
The main result of this paper is that the bound for the discrete case is optimal.

\vskip-0.3cm
\subsection{Formalization as a Markov chain}\label{sec:formalization}
For the sake of simplicity, we let from now on $[a,b]$ denote the discrete interval $\{a,\dots,b\}$ if $a$ and $b$ are integers.
Given a probability distribution $\mu$ on $[1,n]$, the Markov chain
$R=(R_0,R_1,\ldots)$ is defined over the state space $A=[0,n]$ by the transition probabilities
	$$p_{a,a'}=\left\{
	\begin{array}{ll}
	\mu(a-a') & \mbox{ for $a'<a$;}\\
	1- \sum_{1\le d\le a} \mu(d) &  \mbox{ for $a'= a$;}\\
	0  & \mbox{ for $a'>a$}.
	\end{array}
	\right.
	$$
Clearly, $0$ is an absorbing state. 
We define the random variable
\begin{math}
T=\min\{t \mid  R_t=0\}.
\end{math}
Let us write $\E_\mu(T)$ for the expectation 
of $T$ if $R_0$ is uniformly distributed 
in $A-\{0\}=[1,n]$.
We study $\E_\mu(T)$ in dependence on $\mu$. 
In particular, we 
wish to identify distributions $\mu$ that make $\E_\mu(T)$
as small as possible (up to constant factors, where
$n$ is growing).

\label{sec:ClosedExpression}
\begin{observation}\label{obs:mu:one:positive}
If $\mu(1)=0$ then $\E_\mu(T)=\infty.$
\end{observation}
This is because with probability $\frac1n$ position 1 
is chosen as the starting point, and from state 1, the process will never reach 0 if $\mu(1)=0$.
As a consequence, for the whole paper we assume that all 
distributions $\mu$ that are considered satisfy
\begin{equation}\label{eq:obs:mu:one:positive}
	\mu(1) > 0.
\end{equation}

Next we note that it is not hard to derive a ``closed expression'' for $\E_\mu(T)$. Fix $\mu$.
For $a\in A$, let
\begin{math}
F(a)=\mu([1,a])=\sum_{1\le d\le a}\mu(d).
\end{math}
We note recursion formulas for the expected travel time
\begin{math}
  T_a=\E_\mu(\,T \mid R_0=a\,)
\end{math}
when starting from position $a\in A$.
It is not hard to obtain (details are omitted due to space constraints)
\begin{equation}\label{eq:30}
	\E_\mu(T)= \frac{1}{n}\cdot\sum_{1\le a_1 <\cdots<a_\ell\le n}
	    \frac{\mu(a_2-a_1)\cdots\mu(a_{\ell}-a_{\ell-1})}{F(a_1)\cdots F(a_\ell)},
\end{equation}
where the sum ranges over all $2^n-1$ nonempty subintervals $[a_1,a_\ell]$ of $[1,n]$.
By definition of $F(a)$, we see that $\E_\mu(T)$ is a rational function
of $(\mu(1),\ldots,\mu(n))$.
By compactness, there is some $\mu$ that minimizes $\E_\mu(T)$. 
Unfortunately, there does not seem to be an obvious way to use (\ref{eq:30}) to
gain information  about the way $\E_\mu(T)$ depends on $\mu$
or what a distribution $\mu$ that minimizes $\E_\mu(T)$ looks like.

\vskip-0.3cm
\section{Upper bound}
\label{sec:UpperBound}

In this section, we establish upper bounds on $\E_\mu(T)$.
We split the state space $A$ and the set $X$ of possible distances 
into ``orders of magnitude'', arbitrarily choosing $2$ as the base.%
\footnote{$\log$ means ``logarithm to the base $2$'' throughout.}
Let $L=\lfloor\log n\rfloor$, and define $I_i=[2^i,2^{i+1})$, for $0\le i < L$,
and $I_L=[2^L,n]$. 
%
Define 
\begin{displaymath}\label{eq:40}
p_i=\sum_{d\in I_i}\mu(d), \mbox{ for $0\le i \le L$}.
\end{displaymath}
Clearly, then, $p_0+p_1+\cdots+p_L=1$.
To simplify notation, we do not 
exclude terms that mean $p_i$ for $i<0$ or $i>L$.
Such terms are always meant to have value $0$. 
Consider the process $R=(R_0,R_1,\ldots)$.
Assume $t\ge1$ and $i\ge1$.
If $R_{t-1}\ge 2^{i}$ then  
all numbers $d\in I_{i-1}$ will be accepted as steps and 
lead to a progress of at least $2^{i-1}$.
Hence
$$
\Pr(R_{t} \le R_{t-1} - 2^{i-1}\mid R_{t-1}\ge 2^i ) \ge p_{i-1}.
$$
Further, if $R_{t-1}\in I_i$, 
we need to choose step sizes from $I_{i-1}$ at most twice to get below 
$2^{i}$.
Since the expected waiting time for 
the random distances to hit $I_{i-1}$ twice is
$2/p_{i-1}$, the 
expected time process $R$ remains in $I_i$ is not larger than $2/p_{i-1}$.

Adding up over $1\le i \le L$, the expected time process $R$ spends in
the interval $[2,a]$, where $a\in I_j$ is the starting position, is
not larger than
$$
\frac{2}{p_{j-1}}+ \frac{2}{p_{j-2}}+ \ldots + \frac{2}{p_1} +  \frac{2}{p_0}\;.
$$
After the process has left $I_1=[2,3]$, it has reached position $0$ or position $1$,
and the expected time before we hit 0 is not larger than $1/p_0=1/\mu(1)$.
Thus, the expected number $T_a$ of steps to get from $a\in I_j$ to 0 satisfies
\begin{math}
T_a \le \frac{2}{p_{j-1}}+ \frac{2}{p_{j-2}}+ \ldots + \frac{2}{p_1} +  \frac{3}{p_0}\;.
\end{math}
This implies the bound
\begin{displaymath}
\E_\mu(T) \le \frac{2}{p_{L-1}}+  \frac{2}{p_{L-2}}+\ldots + \frac{2}{p_1} + 
\frac{3}{p_0}\;,
\end{displaymath}
for arbitrary $\mu$.
If we arrange that 
\begin{equation}\label{eq:55}
p_0=\cdots=p_{L-1}=\frac1L,
\end{equation}
we will have  $T_a \le (2j+1)L \le (2(\log a)+1)(\log n)=O((\log a)(\log n))=O((\log n)^2)$.
Clearly, then, $\E_\mu(T)=O((\log n)^2)$ as well. 
The simplest distribution $\mu$ with $(\ref{eq:55})$  is
the one that distributes the weight evenly on the powers of 2 below $2^L$:
$$
\mu_{\rm pow2}(d)=\left\{
\begin{array}{ll}
1/L,&\mbox{if $d=2^i$, $0\le i < L$,}\\
0,&	\mbox{otherwise.}
\end{array}
\right.
$$
Thus,
$\E_{\mu_{\scriptstyle\rm pow2}}(T)=O((\log n)^2).$
The ``harmonic distribution'' defined by (\ref{eq:harmonic:distribution}) 
satisfies
$p_i\approx (\ln(2^{i+1})-\ln(2^i))/H_n \approx \ln2/\ln(n) = 1/\log_2 n$,
and we also get 
$T_a =O((\log a)(\log n))$ and 
$\E_{\mu_{\scriptstyle\rm har}}(T)=O((\log n)^2)$.
More generally, all distributions $\mu$ with
$p_0,\ldots,p_{L-1}\ge\alpha/L$, where $\alpha>0$ is constant, 
satisfy $\E_\mu(T)=O((\log n)^2)$.

\vskip-0.3cm
\section{Lower bound}
\label{sec:LowerBound}
We show, as the main result of this paper, 
that the upper bound of Section~\ref{sec:UpperBound} is 
optimal up to a constant factor. 
\begin{theorem}\label{thm:lower:bound}
	$\E_\mu(T)=\Omega((\log n)^2)$ for all distributions $\mu$.
\end{theorem}
This theorem is proved in the remainder of this section.
The distribution $\mu$ is fixed from here on;
we suppress $\mu$ in the notation. 
Recall that we may assume that $\mu(1)>0$. 
We continue to use the intervals $I_0,I_1,I_2,\ldots,I_L$ that partition $[1,n]$,
as well as the probabilities $p_i$, $0\le i \le L$. 

\vskip-0.3cm
\subsection{Intuition}\label{sec:intuition}
The basic idea for the lower bound is the following. 
For the majority of the starting positions, 
the process has to traverse all intervals $I_{L-2},I_{L-3},\ldots,I_1,I_0$.
Consider an interval $I_i$. 
If the process reaches interval $I_{i+1}$,
then afterwards steps of size $2^{i+2}$
and larger are rejected, and so do not help at all for crossing $I_i$.
Steps of size from $I_{i+1}$, $I_i$, $I_{i-1}$, $I_{i-2}$ may be
of significant help.
Smaller step sizes will not help much.
So, very roughly, the expected time to 
traverse interval $I_i$ completely 
when starting in $I_{i+1}$ will be bounded from below by
$$
\frac{1}{p_{i+1} + p_i + p_{i-1} + p_{i-2}},
$$
since  $1/(p_{i+1} + p_i + p_{i-1} + p_{i-2})$ is the waiting
time for the first step with a ``significant'' size to appear. 
If it were the case that there is a constant $\beta>0$ with the property that for each $0\leq i<L-1$ the probability that interval $I_{i+1}$ is visited is at least $\beta$
then it would not be hard to show that the expected travel time is bounded below by
\begin{equation}\label{eq:60}
\sum_{1\le j < L/2}\frac{\beta}{p_{2j+1} + p_{2j} + p_{2j-1} + p_{2j-2}}.
\end{equation}
(We picked out only the even $i=2j$ to avoid double counting.)
Now the sum of the denominators in the sum in (\ref{eq:60}) is at most $2$,
and the sum is minimal when all denominators are equal,
so the sum is bounded below by $\beta\cdot(L/2)\cdot (L/2)/2=\beta \cdot L^2/8$,
hence the expected travel time would be $\Omega(L^2)=\Omega((\log n)^2)$.

It turns out that it is not
straightforward to turn this informal argument into a rigorous proof. 
First, there are (somewhat strange) distributions
$\mu$ for which it is not the case that
each interval is visited with constant probability.
(For example, let $\mu(d)={B^{d-1}\cdot(B-1)/(B^n-1)}$,
for a large base $B$ like $B=n^3$. Then the ``correct''
jump directly to 0 has an overwhelming probability to be chosen first.%
\footnote{The authors thank Uri Feige for pointing this out.})
Even for reasonable distributions $\mu$,
it may happen that some intervals 
or even blocks of intervals are jumped over with high probability.
This means that the analysis of the cost of 
traversing $I_i$ has to take into account
that this traversal might happen
in one big jump starting from
an interval $I_{j}$ with $j$ much larger than $i$.
Second, in a formal argument, 
the contribution of the steps of size smaller than $2^{i-2}$
must be taken into account.

In the remainder of this section, we  give a rigorous proof of the lower bound.
For this, some machinery has to be developed.
The crucial components are a reformulation of 
process $R$ as another process, which as long as possible
defers decisions about what the 
(randomly chosen) starting position is,
and a potential 
function to measure how much progress 
the process has made in direction to its goal,
namely reaching position 0.     

\vskip-0.3cm
\subsection{Reformulation of the process}
\label{sec:ReformulationOfTheProcess}

We change our point of view on the process $R$ (with initial distribution uniform in $[1,n]$).
The idea is that we do not have to fix the
starting position right at the beginning, 
but rather make partial decisions on what
the starting position is as the process advances. 
The information we hold on for step $t$ is 
a random variable $S_t$, with the following interpretation:
if $S_t>0$ then  $R_t$ is uniformly distributed in $[1,S_t]$;
if $S_t=0$ then $R_t=0$.

What properties should the random process $S=(S_0,S_1, \ldots)$ on $[0,n]$ have to be a proper model of the Markov chain $R$ from Section~\ref{sec:formalization}?
We first give an intuitive description of process $S$, and later formally define the corresponding Markov chain.
Clearly, $S_0=n$: the starting position is uniformly distributed in $[1,n]$.
Given $s=S_{t-1}\in[0,n]$, we choose a step length $d$ from $X$,
according to distribution $\mu$. Then there are two cases. 

\emph{Case} 1: $d > s$. ---
If $s\ge1$, this step cannot be used
for any position in $[1,s]$, thus we reject it and let $S_t=s$.
If $s=0$, no further move is possible at all, and we also reject. 

\emph{Case} 2: $d \le s$. --- 
Then $s\ge1$,
and the token is at some position in $[1,s]$.
What happens now depends on the position of the token
relative to $d$, for which we only have a probability distribution. 
We distinguish three subcases: 
\begin{itemize}
\item[(i)] The position of the token is larger than $d$. ---
This happens with probability 
$(s-d)/s$. 
In this case we ``accept'' the step,
and now know that the token is in $[1,s-d]$, uniformly distributed;
thus, we let $S_t=s-d$. 
\item[(ii)] The position of the token equals $d$. ---
This happens with probability $1/s$. 
In this case we ``finish'' the process, and let $S_t=0$.
\item[(iii)]  The position of the token is   
smaller than $d$. ---
This happens with probability 
$
\frac{d-1}{s}.
$
In this case we ``reject'' the step, 
and now know that the token is in $[1,d-1]$, uniformly distributed;
thus, we let $S_t=d-1$. 
\end{itemize}
Clearly, once state 0 is reached,
all further steps are rejected via Case 1.

We formalize this idea  by defining a new Markov chain $S=(S_0,S_1,\ldots)$,
as follows. The state space is $A=[0,n]$.
For a state $s'$, we collect the total probability that 
we get from $s$ to $s'$. If $s'>s$, this probability is 0;
if $s'=s$, this probability is $\sum_{s<d\le n}\mu(d)=1-F(s)$;
if $s'=0$, this probability is $\sum_{1\le d\le s}\mu(d)/s =F(s)/s$;
if $1\le s'< s$, this probability is $(\mu(s'+1) + \mu(s-s'))\cdot s'/s$,
since $d$ could be $s'+1$ or $s-s'$. 
Thus, we have the following transition probabilities:
	$$p_{s,s'}=\left\{
	\begin{array}{ll}
	F(s)/s & \mbox{ if $s > s'= 0$;}\\
	(\mu(s'+1) + \mu(s-s'))\cdot s'/s & \mbox{ if $s > s'\ge1$;}\\
	1 - F(s) & \mbox{ if $s = s'$}.
	\end{array}
	\right.
	$$
Again, several initial distributions are possible for
process $S$.
The version with initial distribution with $\Pr(S_0=n)=1$ is 
meant to describe process $R$. 
Define the stopping time
$$
T_S=\min\{t \mid S_t=0\}.
$$
We note that it is sufficient to analyze process $S$
(with the standard initial distribution).

\begin{lemma}\label{lemma:equivalence}\quad
$
\E(T)=\E(T_S).
$
\end{lemma}


\proof 
For $0\le s \le n$, consider the 
version $R^{(s)}$ of process $R$ 
induced by choosing the uniform distribution 
on $[1,s]$ (for $s\ge1$) resp. $\{0\}$ (for $s=0$)
as the initial distribution. 
We let 
\begin{displaymath}
A^{(s)}=\E(\min\{t \mid R^{(s)}_t=0\}).
\end{displaymath}
Clearly, $A^{(n)}=\E(T)$ and $A^{(0)}=0$.
%
%
We derive a recurrence for $(A^{(0)},\ldots,A^{(n)})$.
Let $s\ge1$, and assume 
the starting point $R_0$ is chosen uniformly at random from $[1,s]$.
We carry out the first step of $R^{(s)}$, which starts with choosing $d$.
The following situations may arise.
\begin{itemize}
	\item[(i)] $d > s$. --- This happens with probability $1-F(s)<1$. 
	This distance will be rejected for all starting points in $[1,s]$, 
	so the expected remaining travel time is $A^{(s)}$ again. 
	\item[(ii)] $1\le d \le s$. For each $d$,
	the probability for this to happen is $\mu(d)$. 
	For the starting point $R_0$ there are three possibilities:
	\begin{itemize}
	 \item[-] $R_0\in[1,d-1]$ (only possible if $d>1$). --- 
	 This happens with probability $\frac{d-1}{s}$. 
	 The remaining expected travel time is
	 $A^{(d-1)}$.
   \item[-] $R_0=d$. --- 
   This happens with probability $\frac{1}{s}$.
   The remaining travel time is $0$.
   \item[-] $R_0\in[d+1,s]$  (only possible if $d<s$). ---
   This happens with probability $\frac{s-d}{s}$.
    The remaining expected travel time in this case is 
   $A^{(s-d)}$.  
   \end{itemize}
\end{itemize}
We obtain:
$$
A^{(s)} = 1 + (1-F(s))A^{(s)} + 
\sum_{1\le d \le s}\!\! \mu(d)\left(\frac{d-1}{s}\cdot A^{(d-1)} + \frac{s-d}{s}\cdot A^{(s-d)}\right).
$$
We rename $d-1$ into $s'$ in the first sum and $s-d$ into $s'$
in the second sum and rearrange to obtain 
\begin{equation}\label{eq:new:42}
A^{(s)} =
\frac{1}{F(s)}\cdot 
\left(1+ \sum_{1\le s' < s}\!\! (\mu(s'+1)+\mu(s-s'))\cdot(s'/s)\cdot A^{(s')}\right).
\end{equation}
Next, we consider process $S$. 
For $0\le s \le n$, let
$S^{(s)}$ be the process obtained from $S$ by choosing $s$ as the starting point.
Clearly, $S^{(0)}$ always sits in $0$,
and $S^{(n)}$ is just $S$. Let
\begin{displaymath}
B^{(s)}=\E(\min\{t\mid S^{(s)}_t =0\}),
\end{displaymath}
the expected number of steps process $S$ needs to reach 0 when starting in $s$.
Then $B^{(0)}=0$ and $B^{(n)}=\E(T_S)$.
We derive a recurrence for $(B^{(0)},\ldots,B^{(n)})$. Let $s\ge1$.
Carry out the first step of $S^{(s)}$, which leads to state $s'$.
The following situations may arise.
\begin{itemize}
	\item[(i)] $s=s'\ge1$. --- This occurs with probability $1-F(s)$,
	and the expected remaining travel time is $B^{(s)}$ again. 
	\item[(ii)] $s'=0$. --- In this case the expected remaining travel time is $B^{(0)}=0$. 
	\item[(iii)] $s>s'\ge1$. ---
	This occurs with probability $(\mu(s'+1)+\mu(s-s'))\cdot s'/s $.
	The expected remaining travel time is $B^{(s')}$.
\end{itemize}
Summing up, we obtain
$$
B^{(s)} = 1 + (1-F(s))B^{(s)} + 
\sum_{1\le s' < s}\!\! (\mu(s'+1)+\mu(s-s'))\cdot(s'/s)\cdot B^{(s')}.$$
Solving for $B^{(s)}$ yields:
\begin{equation}\label{eq:new:50}
B^{(s)} =
\frac{1}{F(s)}\cdot 
\left(1+ \sum_{1\le s' < s}\!\! (\mu(s'+1)+\mu(s-s'))\cdot(s'/s)\cdot B^{(s')}\right).
\end{equation}
Since $A^{(0)}=0=B^{(0)}$ and the recurrences (\ref{eq:new:42}) and (\ref{eq:new:50}) are identical,
we have $\E(T)=A^{(n)}=B^{(n)}=\E(T_S)$, as claimed.
\qed

\subsection{Potential function: Definition and application}%
\label{sec:APotentialFunction}%
%
We introduce a potential function $\Phi$ on the state space $A=[0,n]$ to bound the progress of process $S$.
Our main lemma states that for any $s\in A$, for a random transition from $S_i=s$ to $S_{i+1}$ the expected loss in potential is at most constant (i.e., $\E(\Phi(S_{i+1})-\Phi(S_i) \mid S_{i}=s)=O(1)$).
This implies that $\E(T_S)=\Omega(\Phi(S_0))$.
Since the potential function will satisify $\Phi(S_0)=\Omega(\log^2 n)$, the lower bound follows.

We start by trying to give intuition for the definition. 
A rough approximation to the potential function we use
would be the following: 
For interval $I_i$ there is a term
\begin{equation}\label{eq:90}
\psi_i = \frac{1}{\sum_{0\le j \le L}p_j\cdot c^{|j-i|}},
\end{equation}
for some constant $c$ with $\frac12 < c < 1$, e.\,g., $c=1/\sqrt{2}$.
For later use we note that
\begin{equation}\label{eq:92}
\sum_{1\le i < L}\psi_i^{-1} = \sum_{1\le i < L}{\sum_{0\le j \le L}p_j\cdot c^{|j-i|}}
= \sum_{0\le j \le L} p_j \sum_{1\le i < L} c^{|j-i|} = O(1),
\end{equation}
since $\sum_{0\le j \le L}p_j=1$ and $\sum_{k\ge 0}c^k=\frac{1}{1-c}$.
The term $\psi_i$ tries to give a rough lower bound for the
expected number of steps needed to cross $I_i$ in the following sense: 
The summands $p_j\cdot c^{|j-i|}$ reflect the 
fact that step sizes that are close to $I_i$ will be very helpful for 
crossing $I_i$, and
step sizes far away from $I_i$ might help a little in crossing $I_i$,
but they do so only to a small extent ($j\ll i$) or with small probability ($j\gg i$).
The idea is then to arrange that 
a state $s\in I_k$  has potential about 
\begin{equation}\label{eq:100}
  \Psi_k=\sum_{i\le k} \psi_i. 
\end{equation}
It turns out that analyzing process $S$ on the basis of a potential function
that refers to the intervals $I_i$
is possible but leads to messy calculations and 
numerous cases. 
The calculations become cleaner
if one avoids the use of the intervals in the
definition and in applying the potential function. 
The following definition derives from
  (\ref{eq:90}) and (\ref{eq:100})
 by splitting up the summands $\psi_i$ into contributions from
 all positions $a\in I_i$ and smoothing out the factors 
 $c^{|j-i|}= 2^{|j-i|/2}$, for $a\in I_i$ and $d\in I_j$, into 
 $2^{-|\log a-\log d|/2}$, which is 
 $\sqrt{a/d}$ for $a\le d$ and 
 $\sqrt{d/a}$ for $d\le a$. 
 This leads to the following%
 \footnote{Whenever in the following we use letters $a,b,d$, the range $[1,n]$ is
 implicitly understood.}. 
 Assumption (\ref{eq:obs:mu:one:positive}) 
 guarantees that in the formulas to follow all denominators are nonzero.

\begin{definition}\label{definition:potential}%
For $1\le a \le n$ let
\begin{displaymath}
\sigma_a = \sum_{1\le d\le n}\mu(d)\cdot2^{-|\log a-\log d|/2} =
\sum_{1\le d\le a}\mu(d)\sqrt{\frac{d}{a}} + \sum_{a < d \le n } \mu(d)\sqrt{\frac{a}{d}}
\end{displaymath}
and $\varphi_a = 1/(a\sigma_a)$. 
For $0\le s \le n$ define
\begin{math}
\Phi(s)=\sum_{1 \le a \le s} \phi_a.
\end{math}
The random variable $\Phi_t$, $t=0,1,2,\ldots$, is defined as $\Phi_t=\Phi(S_t)$.
\end{definition}

We note some easy observations and one fundamental fact about $\Phi_t$, $t\ge0$.

\begin{lemma}\label{lemma:potential:lower:bound}\quad
\begin{itemize}\samepage
	\item[{\rm(a)}]  $\Phi_t$, $t\ge0$, is nonincreasing for $t$ increasing.  
	\item[{\rm(b)}]  $\Phi_t=0$  $\Leftrightarrow$ $S_t=0$. 		
	\item[{\rm(c)}]  $\Phi_0 =\Omega((\log n)^2)$ \quad \emph{(}$\Phi_0$ is a number that depends on 
	$n$ and $\mu$\emph{)}.
\end{itemize}
\end{lemma}

\proof (a) is clear since $S_t$, $t\ge0$, is nonincreasing
and the terms $\phi_a$ are positive. --- 
(b) is obvious since $\Phi_t=0$
if and only if 
$\Phi(S_t)$ is the empty sum, which is the case if and only if $S_t=0$. --- We prove (c).
In this proof we use the intervals $I_i$ and the probabilities $p_i$, $0\le i \le L$, 
from Section~\ref{sec:UpperBound}.
We use the notation $i(a)=\lfloor\log a \rfloor=\max\{i\mid 2^i\le a\}$.
We start with finding an upper bound
for $\sigma_a$
by grouping the summands in $\sigma_a$ according to the intervals.
Let $c=1/\sqrt{2}$. 

\begin{align*}
\sigma_a &= \sum_{1\le d\le n}\mu(d)\cdot 2^{-|\log a-\log d|/2}\\
& \le \sum_{j\le i(a)} \;\sum_{d\in I_j} \mu(d) \cdot 2^{ (j + 1 - i(a))/2} +
                          \sum_{j > i(a)} \;\sum_{d\in I_j}  \mu(d) \cdot 2^{ (i(a)+ 1 - j)/2}   \\
&=  \sum_{j\le i(a)} p_j \cdot 2^{(j+1-i(a))/2} +
                          \sum_{j > i(a)} p_j \cdot 2^{(i(a)+1-j)/2}
 =    2c\cdot \left(\; \sum_{0\le j \le L} p_j \cdot c^{|j-i(a)|}  \right) .                   
\end{align*}
Hence
\begin{displaymath}
	\sum_{a\in I_i } \phi_a =  \sum_{a\in I_i } \frac{1}{a\sigma_a}
	\ge
	\frac{2^i}{2c\cdot 2^{i+1} \cdot \left(\; \sum_{0\le j \le L} p_j \cdot c^{|j-i|}   \right)}
	=
	\frac{\psi_i }{4c},
\end{displaymath}
with $\psi_i$ from (\ref{eq:90}).
Thus,
\begin{equation}\label{eq:130}
\Phi_0 
\ge \sum_{0\le i < L} \frac{\psi_i}{4c}.
\end{equation}
Let $u_i=4c/\psi_i$ be the reciprocal of the summand for $i$ in (\ref{eq:130}), $0\le i < L$.
From (\ref{eq:92}) we read off that 
$\sum_{0\le i <L}u_i \le k$, for some constant $k$.
Now $\sum_{0\le i < L} \frac{1}{u_i}$ with $\sum_{0\le i < L} u_i\le k$
is minimal if all $u_i$ are equal to $k/L$. 
Together with (\ref{eq:130}) this entails
$\Phi_0 \ge L\cdot(L/k) =L^2/k=\Omega((\log n)^2)$,
which proves part (c) of Lemma~\ref{lemma:potential:lower:bound}.
\qed

The crucial step in the lower bound proof is to show 
that the progress made by process $S$ in one step,
measured in terms of the potential, is bounded:
\begin{lemma}[\textbf{Main Lemma}]\label{lemma:potential:main}
There is a constant $C$ such that for $0\le s\le n$, we have
\begin{math}
\E(\Phi_{t-1}-\Phi_{t}\mid S_{t-1} = s) \le C.
\end{math}
\end{lemma}
The proof of Lemma~\ref{lemma:potential:main} is the core of the analysis. 
It will be given in Section~\ref{sec:ProofMainLemma}.
To prove Theorem~\ref{thm:lower:bound},
we need the following lemma,
which is stated and proved  (as Lemma 12) in~\cite{Jae07}.
(It is a one-sided variant of Wald's identity.)

\begin{lemma}\label{lem:jaegerskuepper}
Let $X_1,X_2,\ldots{}$ denote random variables with bounded range, 
let $g>0$ and let $T=\min\{t\mid X_1+\cdots+ X_t \ge g\}$. 
If $\E(T)<\infty$ and $\E(X_t \mid T\ge t) \le C$ for all $t\in\N$,
then $\E(T)\ge g/C$.
\end{lemma}
\emph{Proof} of \ref{thm:lower:bound}:
Since $S_t=0$ if and only if $\Phi_t=0$ (Lemma~\ref{lemma:potential:lower:bound}(b)),
the stopping time $T_\Phi=\min\{t\mid \Phi_t=0\}$ of the potential reaching 0 satisfies
$T_\Phi = T_S$.
Thus, to prove 
Theorem~\ref{thm:lower:bound}, 
it is sufficient to show that $\E(T_\Phi) = \Omega((\log n)^2)$.
For this, we let $X_t=\Phi_{t-1}-\Phi_{t}$, the progress made in step $t$
in terms of the potential. 
By Lemma~\ref{lemma:potential:main}, $\E( X_t \mid S_{t-1} = s)\le C$,
for all $s\geq 1$, and hence
$$
\E(X_t \mid T \ge t ) = \E(X_t \mid \Phi(S_{t-1}) > 0) \le C \; .
$$ 
Observe that  $X_1+\cdots+X_t=\Phi_0-\Phi_t$ and hence
$T_\Phi=\min\{t \mid X_1+\cdots+ X_t  \ge \Phi_0\}$.
 Applying Lemma~\ref{lem:jaegerskuepper}, and combining 
 with  Lemma~\ref{lemma:potential:lower:bound}, we get that
$\E(T_\Phi)  \ge \Phi_0/C  = \Omega((\log n)^2)$,
which proves Theorem~\ref{thm:lower:bound}.
\qed 

The only missing part to fill in is the proof of Lemma~\ref{lemma:potential:main}.

\vskip-0.3cm
\subsection{Proof of the Main Lemma (Lemma~\ref{lemma:potential:main})}
\label{sec:ProofMainLemma}

Fix $s\in[1,n]$, and assume $S_{t-1}=s$.
Our aim is to show that
the ``expected potential loss'' is constant, i.\,e., that 
\begin{displaymath}
\E(\Phi_t-\Phi_{t-1}\mid S_{t-1}=s) = O(1).
\end{displaymath}
Clearly,
$
  \E(\Phi_t-\Phi_{t-1}\mid S_{t-1}=s) = 
  \sum_{0\leq x\leq s}\!\Delta(s,x)
$, where
\begin{equation}  
  \Delta(s,x)=\bigl(\Phi(s)-\Phi(x)\bigr)\cdot\Prob(S_t=x\mid S_{t-1}=s). 
\end{equation}
We show that $\sum_{0\leq x\leq s}\Delta(s,x)$ is bounded by a constant, 
by considering $\Delta(s,s)$, $\Delta(s,0)$, and $\sum_{1\leq x<s}\Delta(s,x)$ separately.

For $x=s$, the potential difference $\Phi(s)-\Phi(x)$ is 0, and thus
\begin{equation}\label{eq:160}
  \Delta(s,s)=0.
\end{equation}
\paragraph{\textbf{\boldmath Bounding $\Delta(s,0)$:}}
According to the definition of the process $S$,
a step from $S_{t-1}=s$ to $S_t=0$ has  probability $F(s)/s$.
Since $\Phi(0)=0$, the potential difference is $\Phi(s)$.
Thus, we obtain

$$
  \Delta(s,0)    =
   \frac{1}{s} \cdot \left( \sum_{d\le s}\mu(d)\right) \cdot \left(\sum_{a\le s} \phi_a\right)
  =  \frac{1}{s}\cdot\sum_{a\le s} \;\;  
  \frac{\displaystyle\sum_{d\le s}\mu(d)}{
  {\displaystyle
  \sum_{b\le a}\mu(b)\sqrt{ab} + \sum_{a < b \le n}\! \mu(b)a^{3/2}/\sqrt{b}}}
$$
\begin{align*}
     \leq
   \frac{1}{s}\cdot\sum_{a\leq s}\delta(a),\quad
   \text{where }
   \delta(a)\;=\;
   \frac{\displaystyle\sum_{b\le s}\mu(b)}{
   {\displaystyle
   \sum_{b\le a}\mu(b)\sqrt{ab} + \sum_{a < b \le s}\! \mu(b)a^{3/2}/\sqrt{b}}} \; . 
 \end{align*}

We bound $\delta(a)$.
For $b\le a$ and $\mu(b)\neq 0$, the quotient of the summands in
the numerator and denominator of $\delta(a)$ that correspond to $b$ is $1/\sqrt{ab}\leq \sqrt{a}/a \leq \sqrt{s}/a$.
For $a < b$ and $\mu(b)\neq 0$,
the quotient is
$\sqrt{b}/a^{3/2} \leq \sqrt{s}/a$.
Thus, 
$\delta(a) \leq \frac{\sqrt{s}}{a}$.
This implies (recall that $H_s=\sum_{a\le d\le s}\frac1a$):
 \begin{equation}\label{eq:190}
\Delta(s,0) \le \frac{1}{s}\cdot\sum_{a\le s} \sqrt{s}/a \leq \frac{H_s}{\sqrt{s}}\leq \frac{\ln(s)+1}{\sqrt{s}} < 2.
\end{equation}

\paragraph{\textbf{\boldmath Bounding $\sum_{1\leq x< s}\Delta(s,x)$:}}
Assume $1\leq x<s$.
According to the definition of the process $S$,
\begin{equation*}
  \Prob(S_{t-1}=x\mid S_t=s) = \frac{x}{s}\cdot\bigl(\mu(x+1)+\mu(s-x)\bigr).
\end{equation*}
The potential difference is $\Phi(s)-\Phi(x)=\sum_{x<a\leq s}\phi_a$.
Thus we have
\begin{equation}\label{eq:210}
\sum_{1\leq x<s} \! \Delta(s,x) =
\sum_{1\leq x<s}  \sum_{x<a\le s} \! \phi_a\cdot\frac{x}{s}\cdot\bigl(\mu(x+1)+\mu(s-x)\bigr)=
\frac{1}{s}\cdot\sum_{1<a\leq s}(\lambda_a+\gamma_a),
\end{equation}
where $\lambda_a=\phi_a\cdot\sum_{1\leq x<a}\mu(x+1)\,x$ and $\gamma_a=\phi_a\cdot\sum_{1\leq x<a}\mu(s-x)\,x$.
We bound $\lambda_a$ and $\gamma_a$ separately.
Observe first that
\begin{align}\label{eq:220}
  \lambda_a &=\
  \phi_a\cdot\!\!\!\sum_{2\leq x\leq a}\mu(x)(x-1)\nonumber\\
  &\leq\
  \frac{\displaystyle\sum_{1\leq x\leq a}\mu(x)(x-1)}{
    \displaystyle\sum_{1\leq b\leq a}\mu(b)\cdot \sqrt{a b} + \sum_{a<b\leq n}\mu(b) a^{3/2}/\sqrt{b}
  }   \leq 
  \frac{\displaystyle\sum_{1\leq b\leq a}\mu(b)(b-1)}{
    \displaystyle\sum_{1\leq b\leq a}\mu(b) \sqrt{a b} \;. 
  }
\end{align}
(We used the definition of $\phi_a$, and omitted some summands in the denominator.)
Recall that $\mu(1)>0$, so the denominator is not zero.
For each $b\le a$ we clearly have $\mu(b)(b-1)\le \mu(b)\sqrt{ab}$,
thus the sum in the numerator in (\ref{eq:220}) is smaller 
than the sum in the denominator, and we get $\lambda_a<1$.

Next, we bound $\gamma_a$ for $a\leq s$:
\begin{align*}
  \gamma_a  
  &= 
  \phi_a\cdot\sum_{1\leq x<a}\mu(s-x)\,x
    \ =\
  \phi_a\cdot\sum_{s-a< x< s}\!\mu(x)\,(s-x)\nonumber\\
  &=
  \frac{
    \displaystyle\sum_{s-a< x\leq a}\!\mu(x)(s-x) + \sum_{\max\{a,s-a\}<x<s}\mu(x)(s-x)}{
    \displaystyle\sum_{1\leq b\leq a}\mu(b)\sqrt{a b} + \sum_{a<b\leq n}\mu(b)a^{3/2}/\sqrt{b}
  }\;.
\end{align*}
The denominator is not zero because $\mu(1)>0$.
Hence, if $\mu(x)=0$ for all $s-a<x<s$, then $\gamma_a=0$.
Otherwise, by omitting some of the summands in the denominator we obtain
\begin{equation*}
  \gamma_a \leq
  \frac{
    \displaystyle\sum_{s-a< b\leq a}\mu(b)\,(s-b) + \sum_{\max\{a,s-a\}<b<s}\mu(b)\,(s-b)}{
    \displaystyle\sum_{s-a< b\leq a}\mu(b)\sqrt{a b} + \sum_{\max\{a,s-a\}<b<s}\mu(b)a^{3/2}/\sqrt{b}
  }
\end{equation*}

(If $a\le s/2$, the first sum in both numerator and denominator is empty.)
Now consider the quotient of the summands for each $b$ with $\mu(b) > 0$.
For $s-a<b\leq a$, this quotient is
\begin{equation*}
  \frac{\mu(b)\,(s-b)}{\mu(b)\sqrt{a b}}
  \leq
  \frac{a-1}{\sqrt{a\cdot (s-a+1)}}
  <
  \sqrt{\frac{a}{s-a+1}}
  \leq 
  \sqrt{\frac{s}{s-a+1}}.
\end{equation*}
For $\max\{a,s-a\}<b<s$, the quotient of the corresponding summands is
\begin{equation*}
  \frac{\mu(b) (s-b)}{\mu(b) a^{3/2}/\sqrt{b}}
  \leq 
  \frac{\min\{a,s-a\}\cdot\sqrt{b}}{a^{3/2}}
  \leq 
  \frac{a\cdot\sqrt{s}}{a^{3/2}}
  =\sqrt{\frac{s}{a}}.
\end{equation*}
Hence,
$\gamma_a\leq \sqrt{s/(s-a+1)} + \sqrt{s/a}.$
Plugging this bound on $\gamma_a$ and the bound $\lambda_a<1$ into (\ref{eq:210}), and using that 
\begin{equation*}
\sum_{1\le a \le s}\frac{1}{\sqrt{a}}=1+\sum_{2\le a \le s}\frac{1}{\sqrt{a}}< 1+\int_1^s\frac{dx}{\sqrt{x}}
= 1+[2\sqrt{x}]_1^s=1+2\sqrt{s}-2 < 2\sqrt{s},
\end{equation*}
we obtain
\begin{multline}
  \sum_{1\leq x<s}\!\Delta(s,x)
  <
  \frac{1}{s}\cdot
  \sum_{1< a\leq s}\left(1+\sqrt{\frac{s}{a}}+\sqrt{\frac{s}{s-a+1}} \right)\\
  < 1+ \frac{1}{\sqrt{s}}\left(\sum_{1< a\leq s}\sqrt{\frac{1}{a}}
                   +\sum_{1\leq a< s}\sqrt{\frac{1}{a}}\right)
  <  1+\frac{2}{\sqrt{s}}\sum_{1\leq a\leq s}\frac{1}{\sqrt{a}}
  < 1+\frac{2}{\sqrt{s}}\cdot 2\sqrt{s}
  = 5.\label{eq:270}
\end{multline}
Summing up the bounds from (\ref{eq:160}), (\ref{eq:190}), and (\ref{eq:270}),
we obtain
$$
\E(\Phi_t-\Phi_{t-1}\mid S_{t-1}=s) \le \Delta(s,0) + \sum_{1\leq x<s}\Delta(s,x) + \Delta(s,s)
< 2 + 5 + 0 = 7.
$$
Thus  Lemma~\ref{lemma:potential:main} is proved.%
\qed

\vskip-0.3cm
\section{Open problems} 
1. We conjecture that the method can be adapted to the continuous case 
to prove a lower bound of $\Omega((\log(1/\varepsilon)^2)$
for approximating the minimum of some unimodal function $f$
by a scale-invariant search strategy (see Section~\ref{sec:blind:optimization}). 

2. It is an open problem whether our method can be used 
to prove a lower bound of $\Omega((\log n)^2)$ for finding 
the minimum of an arbitrary unimodal function $f\colon\{0,\ldots,n\}\to\R$
by a scale invariant search strategy.
 
\vskip-0.3cm
\section*{Acknowledgement} 
The authors thank two anonymous referees
for their careful reading of the manuscript and for providing several helpful comments.

\end{document}